\documentclass[pra,aps,showpacs,twocolumn,superscriptaddress,floatfix]{revtex4}
\usepackage{graphicx}
\usepackage{amsmath}
\usepackage{bm} 
\usepackage{color}
\usepackage{amssymb}

\begin{document}

\title{Controlling integrability in a quasi-1D atom-dimer mixture}
\author{D.~S.~Petrov}
\affiliation{Univ. Paris-Sud, CNRS, LPTMS, UMR8626, Orsay, F-91405,
France}
\affiliation{Russian Research Center Kurchatov Institute, Kurchatov Square, 123182 Moscow, Russia}
\author{V. Lebedev}
\author{J.~T.~M.~Walraven}
\affiliation{Van der Waals--Zeeman Institute of the University of Amsterdam, Science Park 904, 1098 XH Amsterdam, The Netherlands}

\date{\today}

\begin{abstract}
We analytically study the atom-dimer scattering problem in the near-integrable limit when the oscillator length $l_0$ of the transverse confinement is smaller than the dimer size, $\sim l_0^2/|a|$, where $a<0$ is the interatomic scattering length. The leading contributions to the atom-diatom reflection and break-up probabilities are proportional to $a^6$ in the bosonic case and to $a^8$ for the $\uparrow$-$\uparrow\downarrow$ scattering in a two-component fermionic mixture. We show that by tuning $a$ and $l_0$ one can control the ``degree of integrability'' in a quasi-1D atom-dimer mixture in an extremely wide range leaving thermodynamic quantities unchanged. We find that the relaxation to deeply bound states in the fermionic (bosonic) case is slower (faster) than transitions between different Bethe ansatz states. We propose a realistic experiment for detailed studies of the crossover from integrable to nonintegrable dynamics.
\end{abstract}
\maketitle

\section{Introduction}
\label{sec:intro}

Ultracold gases allow unprecedented experimental control over key parameters of a many-body system \cite{Bloch-Dalibard-Zwerger} and offer unique opportunities to explore and understand its out-of-equilibrium behavior. Particularly interesting are one-dimensional systems which can be more or less closely approximated by integrable models and, therefore, allow one to study the effects of integrability and deviations from it. One-dimensional bosons \cite{1DBosons} and fermions \cite{CI-molecule,Randy1DFermions} are, in fact, quasi-1D systems in which the transverse motion is frozen out by a very tight confinement. By integrating out the transverse motion in a two-body problem one obtains a 1D delta-function potential for the effective interatomic interaction \cite{Olshanii}. In this manner quasi-1D spinless bosons and spin-1/2 fermions are modelled by the integrable Lieb-Liniger \cite{Lieb-Liniger} and Yang-Gaudin \cite{CNYang,Gaudin} hamiltonians respectively.

Quantum integrability implies that the $N$-body scattering is nondiffractive, i.e. the outgoing momenta of particles are restricted to be only rearrangements of the incoming ones \cite{Sutherland}. The two-body scattering in 1D is necessarily diffractionless for equal mass particles. However, already for $N=3$ the nondiffractive rule has nontrivial consequences. In particular, the probability of reflection and break-up in atom-dimer collisions vanishes independent of the collision energy \cite{McGuire1964}. This manifestly quantum phenomenon, which has never been observed experimentally, is a microscopic analog of the reflectionless scattering of solitons.

Small imperfections of a realistic quasi-1D system compared to its idealized integrable counterpart are practically irrelevant as long as we are interested, for example, in the equation of state or thermodynamic quantities. However, the statistics of energy levels, localization of eigenstates, decay of excitations, thermalization, response to an external perturbation, transport, and other dynamical properties are sensitive to deviations from integrability. Muryshev {\it et al.} \cite{Muryshev2002} have found that virtual excitations of transverse modes are responsible for the dissipative dynamics of dark solitons in a weakly interacting quasi-1D Bose gas. They have shown (see also \cite{Sinha,Mazets2008}) that the effect of the transverse degrees of freedom can be accounted for by adding a local three-body interaction term into the nonlinear Schr\"odinger equation (equivalent in this case to the Lieb-Liniger model). Yurovsky {\it et al.} \cite{Yurovsky} have numerically solved a purely 1D three-boson problem near a Feshbach resonance and found that the integrability is lifted when the two-body coupling constant becomes energy-dependent.

In this paper we solve the quasi-1D atom-dimer scattering problem by extending the approach of Mora {\it et al.} \cite{MoraL,MoraA} to the case of finite collision momenta. In the nearly-integrable limit, $|a|/l_0\ll 1$, we analytically calculate the scattering amplitudes and find that at a finite collision momentum $q$ the leading contributions to the reflection and break-up probabilities are proportional to $a^8/l_0^{10}q^2$ for fermions and to $a^6/l_0^8q^2$ for bosons. We show that the diffraction in the fermionic case originates mainly from the effective range corrections to the quasi-1D two-body scattering amplitude. In contrast, the bosonic result is due to a diffractive scattering of the atom by the transversely excited part of the dimer, consistent with the local three-body potential of Refs.~\cite{Muryshev2002,Sinha,Mazets2008}. 

The strong $a$-dependence of diffraction suggests a way of creating a system with fixed thermodynamic properties but with a tunable integrability parameter, the fermionic case being more practical due to the higher power in the $a$-dependence and, as we also discuss, due to the suppression of inelastic processes. A four-fold increase of $a$ and doubling of $l_0$ does not change the effective 1D model (the effective coupling constant is unchanged), whereas the reflection probability in atom-dimer collisions increases by a factor of 64 in the fermionic case and by a factor of 16 for bosons. This effect can be verified by colliding clouds of atoms and dimers in a similar but more tunable fashion as the quantum Newton's cradle experiment of Kinoshita {\it et al.} \cite{Kinoshita}. We point out that a quasi-1D two-component $^{40}$K mixture close to a zero crossing for the interspecies scattering length is a promising candidate for exploring this and other phenomena that are sensitive to deviations from integrability. 

The paper is organized as follows. In Sec.~\ref{sec:STM} we introduce our notation and briefly rederive the three-body integral equations of Refs.~\cite{MoraL,MoraA}. In Sec.~\ref{sec:analytic} we separate the ``integrable'' part from the ``perturbation'' and develop the corresponding perturbation theory by constructing the resolvent of the integrable part, the small parameter being the ratio of the dimer binding energy to the confinement frequency. Then, we present the perturbative results for the reflection, transmission, and break-up probabilities as well as the three-body recombination rate constant in the fermionic (Sec.~\ref{sec:measurable}) and bosonic (Sec.~\ref{sec:bosons}) cases. The relaxation and recombination to deep molecular states are discussed in Sec.~\ref{sec:relaxation}. In Sec.~\ref{sec:discussion} we propose an experiment in which one can control the degree of integrability and discuss effects that can be studied there from a more general perspective. We conclude in Sec.~\ref{sec:conclusions}.

\section{The three-body integral equation}
\label{sec:STM}

Let us first consider the fermionic $\uparrow$-$\uparrow$-$\downarrow$ system in a cylindrical harmonic trap. We use units $\hbar=m=\omega=1$, where $m$ is the particle mass and $\omega$ is the frequency of the radial confinement. The oscillator length $l_0=\sqrt{\hbar/m\omega}=1$ is our unit of length. After separating the center of mass motion the noninteracting three-body Hamiltonian reads:
\begin{equation}\label{Hamiltonian}
H_0=-\nabla_{{\bf r}_1}^2-\nabla_{{\bf r}_2}^2+{\mbox{\boldmath$\rho$}}_1^2/4+{\mbox{\boldmath$\rho$}}_2^2/4-2,
\end{equation}
where ${\bf r}_1=\{x_1,{\mbox{\boldmath$\rho$}}_1\}$ is the distance from one of the $\uparrow$-atoms to the $\downarrow$-atom, and $\sqrt{3}{\bf r}_2/2=\{\sqrt{3}x_2/2,\sqrt{3}{\mbox{\boldmath$\rho$}}_2/2\}$ is the distance from their center of mass to the second $\uparrow$-atom. The last term in Eq.~(\ref{Hamiltonian}) shifts the ground state energy to zero. Including interactions the Schr\"odinger equation reads:
\begin{equation}\label{Schr}
(H_0-E)\Psi({\bf r}_1,{\bf r}_2)=-[U(r_1)+U(|{\bf r}_1-\sqrt{3}{\bf r}_2|/2)]\Psi({\bf r}_1,{\bf r}_2),
\end{equation} 
where the interspecies interaction is taken as the zero-range Fermi pseudopotential
\begin{equation}\label{pseudopot}
U(r) \cdot = 4\pi a \delta ({\bf r})\partial (r \cdot)/\partial r.
\end{equation}

The wavefunction of the system should be antisymmetric with respect to the permutation of identical $\uparrow$-fermions, i.e. 
\begin{equation}\label{symmetry}
\Psi({\bf r}_1,{\bf r}_2)=-\Psi({\bf r}_1\rightarrow \tilde{\bf r}_1,{\bf r}_2\rightarrow \tilde{\bf r}_2),
\end{equation}
where
\begin{eqnarray}\label{coordchange}
\tilde{\bf r}_1&=& \tilde{\bf r}_1 ({\bf r}_1,{\bf r}_2)=({\bf r}_1-\sqrt{3}{\bf r}_2)/2,\nonumber \\
\tilde{\bf r}_2&=& \tilde{\bf r}_2 ({\bf r}_1,{\bf r}_2)=-(\sqrt{3}{\bf r}_1+{\bf r}_2)/2.
\end{eqnarray}

We now introduce an auxiliary function $f$ proportional to the regular part of $\Psi$ in the vicinity of $r_1 = 0$:
\begin{equation}\label{f}
\lim_{r_1\rightarrow 0} \partial[r_1\Psi({\bf r}_1,{\bf r}_2)]/\partial r_1=-f({\bf r}_2)/4\pi a.
\end{equation}
The function $f$ can be considered as the wavefunction for the atom-dimer relative motion. In particular, the atom-dimer scattering phase shifts can be read off of its long-distance asymptote. In our problem the atom and the dimer are initially in the transverse ground state and, since the total radial angular momentum is conserved, it is sufficient to consider $f(x,{\mbox{\boldmath$\rho$}})=f(x,\rho)$. In the rest of this section we will derive an equation for this function and discuss its relation to the scattering amplitudes.

Comparing Eqs.~(\ref{pseudopot}) and (\ref{f}) we rewrite Eq.~(\ref{Schr}) in the form
\begin{equation}\label{Schrnew}
(H_0-E)\Psi({\bf r}_1,{\bf r}_2)=f({\bf r}_2)\delta({\bf r}_1)-f(\tilde{\bf r}_2)\delta(\tilde{\bf r}_1).
\end{equation} 
Then let us expand $f$ in eigenstates of a single-particle quasi-1D Hamiltonian:
\begin{equation}\label{fexpanded}
f(x,\rho)=\sum_{m=0}^{\infty}\int\frac{dk}{2\pi}f_m(k)R_m(\rho) e^{ikx},
\end{equation}
where $R_m(\rho)=L_m(\rho^2/2)\exp(-\rho^2/4)/\sqrt{2\pi}$ are normalized radially symmetric eigenfunctions of a 2D harmonic oscillator,
\begin{equation}\label{2Doscillator}
(-\nabla_{\mbox{\boldmath$\rho$}}^2+\rho^2/4-1)R_m(\rho)=2mR_m(\rho),
\end{equation}
and $L_m$ are Laguerre polynomials. The solution of Eq.~(\ref{Schrnew}) can now be written as
\begin{eqnarray}\label{Psithroughf}
\Psi({\bf r}_1,{\bf r}_2)&=&\sum_{m=0}^{\infty}\int\frac{dk}{2\pi}f_m(k)\left[R_m(\rho_2) e^{ikx_2}G_{E-k^2-2m}({\bf r}_1)\right.\nonumber\\
&-&\left.R_m(\tilde{\rho}_2) e^{ik\tilde{x}_2}G_{E-k^2-2m}(\tilde{\bf r}_1)\right].
\end{eqnarray}
The Green function $G_E({\bf r})$ satisfies the equation $(-\nabla_{\bf r}^2+\rho^2/4-1-E)G_E({\bf r})=\delta({\bf r})$ and can be explicitly written as
\begin{eqnarray}\label{Green}
G_E({\bf r})&=&\int_0^\infty \frac{\exp[-(\rho^2/4) \coth\tau-x^2/4\tau+E\tau+\tau]}{(4\pi)^{3/2}\sqrt{\tau}\sinh\tau} d\tau\nonumber\\
&=&\frac{1}{4\pi r}+\frac{\zeta(1/2,-E/2)}{4\pi\sqrt{2}}+o(r),
\end{eqnarray}
where we also present the two leading terms in its small-$r$ expansion. In Eq.~(\ref{Green}) $\zeta$ is the Hurwitz zeta function.

We now substitute (\ref{Psithroughf}) into the left hand side of Eq.~(\ref{f}), multiply the resulting equation by $R_n(\rho_2)\exp(-ipx_2)$, and integrate it over ${\bf r}_2$. The result is
\begin{equation}\label{Thefequation}
\frac{\zeta(1/2,p^2/2-E/2+n)}{4\pi\sqrt{2}}f_n(p)-\sum_{m=0}^{\infty}\hat{M}_{nm}f_m(p)=-\frac{f_n(p)}{4\pi a},
\end{equation}
where the operator $\hat{M}_{nm}$ is defined as
\begin{equation}\label{hatM}
\hat{M}_{nm}f_m(p)=\int\frac{dk}{2\pi}M_{nm}[E-(4/3)(p^2+k^2+pk)]f_m(k)
\end{equation}
with the kernel
\begin{eqnarray}\label{M}
&&\hspace{-1.5cm}M_{nm}(E)=(1/\sqrt{3}\pi)\int_0^\infty\int_0^\infty L_n(z)L_m(z/4)\nonumber\\
&&\hspace{-1cm}\times \exp\left[-\frac{5+3\coth\tau}{8}z\right] dz \frac{\exp[(E-2m)\tau]}{1-\exp (-2\tau)}d\tau. 
\end{eqnarray}
The function $f_0(p)$ can be thought of as the atom-dimer scattering wavefunction in momentum space and its singularities can be made explicit:  
\begin{equation}\label{fthroughg}
f_0(p)=2\pi \delta(p-q)-i g_0(p)2q/(p^2-q^2-i0),
\end{equation}
where the scattering momentum $q>0$ satisfies
\begin{equation}\label{BindingEnergy}
\zeta(1/2,q^2/2-E/2)=-\sqrt{2}/a=\zeta(1/2,E_d/2).
\end{equation}  
Here we introduce the dimer binding energy $E_d>0$ which equals $q^2-E$ and, therefore, the second equality in Eq.~(\ref{BindingEnergy}) directly relates $a$ and $E_d$. In the case of small and negative $a$ the binding energy is also small. Expanding the Hurwitz zeta function [see Eq.~(\ref{Hurwitz})] we obtain $E_d\approx a^2\ll 1$.

The function $g_0(p)$ is smooth, and its values at $p=q$ and $p=-q$ are related to the atom-dimer transmission and reflection amplitudes:
\begin{equation}\label{amplitudes}
t(q)=1+g_0(q),\hspace{0.5cm}r(q)=g_0(-q).
\end{equation}

The functions $f_n$ for $n>0$ also have poles (on the real axis) if the collision energy is high enough to excite the relative atom-dimer motion to the $n$-th transverse state, i.e. $q^2>2n$. The correct rule of integrating the poles can be enforced by the ansatz $f_n(p)=-2i g_n(p) q_n/(p^2-q_n^2-i0)$, where $q_n=\sqrt{q^2-2n}$ and the functions $g_n(p)$ are smooth. The amplitudes of the forward and backward propagating waves in the $n$-th channel are given by $g_n(q_n)$ and $g_n(-q_n)$ respectively.

The break-up channel opens for $q^2>E_d$, i.e. for positive $E$. We can still use Eq.~(\ref{Psithroughf}) in this case by choosing the retarded Green function $G_{E>0}$ which requires that the flux of the released atoms be directed to infinity in the plane $\{x_1,x_2\}$. The retarded Green function is obtained by the analytic continuation of $G_{E<0}$ to positive energies along a contour in the upper halfplane, or, equivalently, by substituting $E\rightarrow E+i0$. One should then proceed with solving Eq.~(\ref{Thefequation}) respecting the branch cuts and the rules of residue integrations. For example, $\sqrt{-E}$ should be substituted by $-i\sqrt{E}$ for $E>0$. The break-up probability can then be calculated, for example, from the equation
\begin{equation}\label{BreakUpProb}
P_{b}(q)=1-|t(q)|^2-|r(q)|^2-\sum_{n=1}^{[q^2/2]}|g_n(q_n)|^2+|g_n(-q_n)|^2,
\end{equation} 
which follows from the atom number conservation law.

\section{Perturbative formalism}
\label{sec:analytic}

Mora {\it et al.} \cite{MoraL,MoraA} have derived Eq.~(\ref{Thefequation}), projected it to the lowest transverse channel by setting $f_{n>0}\equiv 0$, and solved it numerically at zero collision energy. In this manner they determine the 1D scattering lengths for the even and odd channels, $a_{ad}$ and $b_{ad}$, as functions of $E_d$. They also note that in the limit $E_d\rightarrow 0$ Eq.~(\ref{Thefequation}) takes the form of a purely 1D integral equation, the solution of which can be found analytically from the Bethe ansatz. The transmission and reflection amplitudes are known in this case for any collision energy:
\begin{equation}\label{t0r0}
t^{(0)}(q)=\frac{-\sqrt{q^2-E}+\sqrt{3}iq}{\sqrt{q^2-E}+\sqrt{3}iq},\hspace{0.5cm}r^{(0)}(q)\equiv 0,
\end{equation}
and the break-up reaction probability also strictly vanishes.

In order to derive the corrections to the amplitudes (\ref{t0r0}) we return to Eq.~(\ref{Thefequation}) and separate out its integrable part in the following manner. We represent the Hurwitz zeta function as
\begin{equation}\label{Hurwitz}
\zeta\left(\frac{1}{2},\frac{-E}{2}\right)=\sqrt{\frac{2}{-E}}+\int_0^\infty \left[\frac{2\exp(E\tau)}{\exp(2\tau)-1}-\frac{1}{\tau}\right]\frac{d\tau}{\sqrt{2\pi\tau}},
\end{equation}
where we make explicit the part diverging at $E\rightarrow 0$. Similarly, Eq.~(\ref{M}) for the ground transverse channels reads
\begin{equation}\label{M00}
M_{00}(E)=\frac{1}{\pi\sqrt{3}}\left(\frac{1}{-E}+\int_0^\infty \frac{\exp(E\tau)d\tau}{4\exp(2\tau)-1}\right).
\end{equation}
Now, by using Eqs.~(\ref{BindingEnergy}), (\ref{Hurwitz}), and (\ref{M00}) we rewrite the $n=0$ part of Eq.~(\ref{Thefequation}) as
\begin{equation}\label{TheOperEq}
(\hat{L}-\lambda_q)f_0(p)=V(p)f_0(p)+\hat{M}'f_0(p)+\sum_{m>0}\hat{M}_{0m}f_m(p),
\end{equation}
where the operator $\hat{L}$ is defined as
\begin{equation}\label{L}
\hat{L}f(p)=\frac{f(p)}{4\pi\sqrt{p^2-E}}-\frac{\sqrt{3}}{4\pi}\int\frac{dk}{2\pi}\frac{f(k)}{k^2+p^2+kp-3E/4},
\end{equation}
$\lambda_q=1/(4\pi\sqrt{q^2-E})$, the function $V(p)$ equals
\begin{equation}\label{V}
V(p)=\frac{1}{4\pi}\int_0^\infty \frac{\exp(E\tau-q^2\tau)-\exp(E\tau-p^2\tau)}{\exp(2\tau)-1}\frac{d\tau}{\sqrt{\pi\tau}},
\end{equation}
and $\hat{M}'$ is an operator defined by
\begin{equation}\label{Mprime}
\hat{M}'f(p)=\frac{1}{\sqrt{3}\pi}\int\frac{dk}{2\pi}f(k)\int_0^\infty d\tau\frac{e^{E\tau-(4/3)(p^2+k^2+pk)\tau}}{4e^{2\tau}-1}.
\end{equation}

In the limit $E_d\rightarrow 0$ the right hand side of Eq.~(\ref{TheOperEq}) vanishes. Neglecting it one arrives at an integral equation describing the purely one-dimensional integrable case \cite{Dodd,MoraA} which is our zeroth order starting point. Since the atom-dimer scattering solution is known from the Bethe ansatz for any momentum $\tilde{q}\in (-\infty,\infty)$, the operator $\hat{L}$ is easily diagonalized. Namely, the eigenstate equation
\begin{equation}\label{TheZeroEq}
\hat{L}\chi_{\tilde{q}}(p)=\lambda_{\tilde{q}}\chi_{\tilde{q}}(p)
\end{equation}
is solved by
\begin{equation}\label{chi}
\chi_{\tilde{q}}(p) = 2\pi\delta(p-\tilde{q})-i\frac{t^{(0)}(\tilde{q})-1}{p-\tilde{q}-i0}+\frac{i[t^{(0)}(\tilde{q})-1](p+2\tilde{q})}{p^2+\tilde{q}^2+p\tilde{q}-3E/4}.
\end{equation}
Accordingly, as the zeroth order solution of Eq.~(\ref{TheOperEq}) we take
\begin{equation}\label{fzero}
f_0^{(0)}=\chi_q,\,f_{m>0}^{(0)}\equiv 0.
\end{equation}

Consider for simplicity that the collision energy is smaller or of order $E_d$, so that there is a single small parameter $E_d\ll 1$. We then substitute (\ref{fzero}) into Eq.~(\ref{TheOperEq}) and estimate the magnitude of different terms by using the fact that typical momenta involved in $\chi_q(p)$ are of order $\sqrt{E_d}$. We see that $\hat{L}\sim\lambda_q\sim 1/\sqrt{E_d}$. Expanding Eq.~(\ref{V}) to the leading order in $E$, $q$, and $p$ we get
\begin{equation}\label{Vexpanded}
V(p)\approx [\zeta(3/2)/16\sqrt{2}\pi] (p^2-q^2),
\end{equation}
where $\zeta(3/2)\approx 2.61$ is the Riemann zeta function. From Eq.~(\ref{Vexpanded}) we see that $V\sim E_d$ when acting on $\chi_q(p)$ and the leading order correction to $f_0$, which can be written as $(\hat{L}-\lambda_q)^{-1}V(p)\chi_q(p)$, is of order $E_d^{3/2}\chi_q$.

To understand the physics behind the $V$-term in Eq.~(\ref{TheOperEq}) imagine a purely-1D problem in which $\uparrow$ and $\downarrow$ fermions interact via a delta-function potential, the strength of which depends on their collision energy. If this dependence is chosen such that the corresponding scattering amplitude matches the one obtained for two quasi-1D atoms \cite{Bergeman}, we arrive exactly at the three-body equation $(\hat{L}-\lambda_q)f(p)=V(p)f(p)$. The $V$-term then reflects the effective range corrections to the two-body interaction due to virtual transverse excitations. That this type of perturbation breaks integrability in a system of three identical 1D bosons has been shown numerically by Yurovsky {\it et al.} \cite{Yurovsky}.

As a side remark we note that so far we have been considering the idealized zero-range 3D interaction potential (\ref{pseudopot}). Corrections corresponding to the finite range of a realistic 3D potential or to a finite Feshbach resonance width can be incorporated into the formalism of Sec.~\ref{sec:STM} by introducing an energy dependent 3D scattering length, $1/a\rightarrow 1/a(\epsilon_{coll})$. The collision energy $\epsilon_{coll}$ is defined as the kinetic energy, $-\nabla_{{\bf r}_1}^2$, of the relative motion of two atoms when they are close to each other but still outside of the support of the 3D interaction potential. If their relative motion with respect to the third atom is described by the wavefunction $R_m(\rho_2)e^{ikx_2}$ one can see from Eq.~(\ref{Hamiltonian}) that $\epsilon_{coll}=E-k^2+1-2m$. Adding the effective range term, $1/a\rightarrow 1/a-(r_0/2)\epsilon_{coll}$, one arrives at Eq.~(\ref{Thefequation}) where  
\begin{equation}\label{astar}
1/a\rightarrow 1/a-(r_0/2)(E-p^2+1-2n).
\end{equation} 
The right hand side of Eq.~(\ref{TheOperEq}) then acquires an additional term
\begin{equation}\label{Vstar}
\tilde{V}(p)f_0(p) = -(r_0/8\pi)(p^2-q^2)f_0(p),
\end{equation}
which has the same form as Eq.~(\ref{Vexpanded}). Therefore, if we know the correction to $f$ due to the perturbation $V$, the perturbation $\tilde{V}$ does not require any special treatment. We simply multiply the $V$-result by $1-2\sqrt{2}r_0/\zeta(3/2)$. Note that $\tilde{V}\sim V$ only when $r_0$ is comparable to the transverse oscillator length, which is typically of order 100~nm and much larger than the physical range of van der Waals potentials for neutral atoms. Therefore, one should care about $\tilde{V}$ only when $r_0$ is anomalously large, in particular, in the case of a very narrow Feshbach resonance.

Let us now turn to the second term in the right hand side of Eq.~(\ref{TheOperEq}). Approximating $e^{E\tau-(4/3)(p^2+k^2+pk)\tau}$ by 1 in Eq.~(\ref{Mprime}) we obtain $\hat{M}'f(p)\approx \log(2/\sqrt{3})/\sqrt{3}\pi  \int f(p) dp/2\pi$, i.e. this term is local in the position representation. Adopting the confinement-induced resonance terminology the $M'$-term describes the interaction of the third atom with the ``closed channel'' molecule formed by the first two atoms. If we were dealing with bosons (see Sec.~\ref{sec:bosons}) the $M'$-term would give a correction to $f_0$ of order $E_d\chi_q$, i.e. it would be more important than the $V$-term. However, for the fermionic $\uparrow$-$\uparrow$-$\downarrow$ system $\int \chi_q(p) dp=0$ which follows from the fact that three atoms can not be at one point in space. In this case the leading contribution to the $M'$-term is obtained by further expanding Eq.~(\ref{Mprime}) and corresponds to the odd-channel interaction of the ``closed channel'' molecule with an atom. We will quantify this interaction later [see Eq.~(\ref{Mprimematrix})]. Now it is sufficient for us to say that $M'\sim E_d^{3/2}$ when acting on $\chi_q(p)$ and is thus less important than the $V$-term.

Finally, in order to estimate the contribution of the last term in Eq.~(\ref{TheOperEq}) we have to consider Eq.~(\ref{Thefequation}) for $n>0$. To the leading order we obtain
\begin{equation}\label{fmleadingorder}
f_n(p)\approx -(1/\lambda_q)\hat{M}_{n0}\chi_q(p).
\end{equation}
From Eq.~(\ref{hatM}) it follows that $M_{n0}(E\rightarrow 0)={\rm const}$ and similarly to the operator $\hat{M}'$ the order of magnitude of $f_n$ depends on whether $\int \chi_q(p) dp$ vanishes or not. For fermions it does and we have $f_n\sim E_d^2\chi_q$. Substituting this result into the last term in Eq.~(\ref{TheOperEq}) we see that the contribution of the higher transverse channels can be safely neglected even compared to the $M'$-term (cf. \cite{MoraA}). This is what we will do and until further notice we omit the subscript 0 of the function $f_0(p)$.

In the first approximation Eq.~(\ref{TheOperEq}) can be solved by substituting $f^{(0)}$ in its right hand side and by inverting the operator $\hat{L}-\lambda_q$. We keep the operator $\hat{M}'$ in play because its leading order contribution is still larger than higher order terms related to $V$. Accordingly, we call $f^{(1)}=(\hat{L}-\lambda_q)^{-1}(V+\hat{M}')f^{(0)}$ the first order correction to $f$ although it actually contains the leading order term proportional to $E_d^{3/2}\chi_q$ as well as the next-to-leading term $\propto E_d^2\chi_q$. To diagonalize the operator $\hat{L}$ we expand $f(p)$ in the basis of its eigenstates $\chi_q$ [see Eq.~(\ref{TheZeroEq})]:
\begin{equation}\label{fexpansion}
f(p)=\int \alpha_k \chi_k(p) dk/2\pi.
\end{equation}
One can directly show that these eigenfunctions are orthonormal in the sense 
\begin{equation}\label{scalarproduct}
\langle q_1|q_2\rangle=\int_{-\infty}^\infty\bar{\chi}_{q_1}(p)\chi_{q_2}(p)dp/2\pi=2\pi\delta(q_1-q_2).
\end{equation}
Here $\bar{\chi}_q=\chi_q^*$ if $E<q^2$, which is always the case below the break-up threshold. Otherwise, as mentioned in the end of Sec.~\ref{sec:STM} we require all the functions that we are dealing with be analytic in the upper halfplane of complex variable $E$. In particular, the functions $\chi_q$ and $\bar{\chi}_q$ are obtained by analytic continuation of (\ref{chi}) and its complex conjugate from the $E<q^2$ to $E>q^2$ part of the real axis following a path in the upper halfplane. 

Changing the basis from plane waves to $\chi_q$ diagonalizes the left hand side of Eq.~(\ref{TheOperEq}):
\begin{equation}\label{Eqalpha}
(\lambda_k - \lambda_q)\alpha_k=\int \langle k|V+\hat{M}'| k'\rangle \alpha_{k'} dk'/2\pi.
\end{equation}
The matrix elements in Eq.~(\ref{Eqalpha}) are
\begin{eqnarray}\label{Vmatrix}
\langle k|V| k'\rangle &\approx& \frac{C}{4\pi} \frac{(k\sqrt{k'^2-E}-k'\sqrt{k^2-E})/(k-k')}{4(k^2+k'^2+kk')/3-E}\nonumber\\
&&\hspace{-1cm}\times(\sqrt{k^2-E}+i\sqrt{3}k)(\sqrt{k'^2-E}-i\sqrt{3}k')
\end{eqnarray}
and
\begin{equation}\label{Mprimematrix}
\langle k|\hat{M}'| k'\rangle \approx \frac{C'}{4\pi} (i\sqrt{3}k+\sqrt{k^2-E})(i\sqrt{3}k'-\sqrt{k'^2-E}),
\end{equation}
where
\begin{eqnarray}\label{C}
C&=&\sqrt{3}\zeta(3/2)/8\sqrt{2}\approx 0.400,\nonumber\\
C'&=&{\rm Li}_2(1/4)/\sqrt{3}\approx 0.155,\nonumber
\end{eqnarray}
and ${\rm Li}_2(1/4)$ is the polylogarithm function. When calculating the matrix element (\ref{Vmatrix}) we use the approximation (\ref{Vexpanded}) and in Eq.~(\ref{Mprimematrix}) we also retain only the first nonvanishing term in the expansion of Eq.~(\ref{Mprime}). Equations~(\ref{Eqalpha}-\ref{Mprimematrix}) are therefore valid for $q^2,k^2,k'^2,E\ll 1$. 

The iterative solution of Eq.~(\ref{Eqalpha}) is now straightforward. We substitute the zeroth order term $\alpha_k^{(0)}=2\pi\delta(k-q)$ into its right hand side and obtain 
\begin{equation}\label{alpha1}
\alpha_k^{(1)}=\langle k|V+\hat{M}'| q\rangle /(\lambda_k - \lambda_q).
\end{equation}
We then find $f^{(1)}(p)$ from Eq.~(\ref{fexpansion}). The corresponding corrections to the reflection and transmission amplitudes (\ref{t0r0}) can be calculated from the residues of $f^{(1)}(p)$ at $p=-q$ and $p=q$, respectively [see Eqs.~(\ref{fthroughg}) and (\ref{amplitudes})]. These residues are obtained by performing integration in Eq.~(\ref{fexpansion}) in the vicinities of $k=\mp q$. Close to these points the matrix elements (\ref{Vmatrix}) and (\ref{Mprimematrix}) are smooth and $1/(\lambda_k - \lambda_q)\approx -8\pi(q^2-E)^{3/2}/(k^2-q^2-i0)$. Here the positions of the poles with respect to the real axis are chosen such that there is no incoming wave with momentum $-q$. The corresponding correction to the transmission amplitude reads
\begin{eqnarray}\label{t1result}
&&\hspace{-1cm}t^{(1)}(q)/t^{(0)}(q)=-4\pi i (q^2-E)^{3/2}\langle q|V+\hat{M}'| q\rangle /q\nonumber\\
&&\hspace{-1cm}=-i E_d^2 \frac{C(1-q^2/E_d)-C'E_d^{1/2}(1+3q^2/E_d)}{q},
\end{eqnarray}
and for the reflection amplitude we get
\begin{eqnarray}\label{r1result}
&&\hspace{-1cm}r^{(1)}(q)=-4\pi i (q^2-E)^{3/2}\langle -q|V+\hat{M}'| q\rangle /q\nonumber\\
&&\hspace{-1cm}=-i E_d^2 \frac{C/(1+q^2/3E_d)-C'E_d^{1/2}}{q}(1-i\sqrt{3q^2/E_d})^2,
\end{eqnarray}
Equations~(\ref{t1result}) and (\ref{r1result}) give the first two leading corrections ($\propto E_d^2$ and $\propto E_d^{5/2}$) to the atom-dimer transmission and reflection amplitudes for small $E_d$. The validity of these equations requires that the scattering momentum be in the interval $E_d^2 \ll q \ll 1$. In particular, the atom-dimer collision energy $q^2$ can be above or below the break-up threshold, $E_d$. The unphysical divergence of $t^{(1)}(q)$ and $r^{(1)}(q)$ at very small momenta $q\lesssim E_d^2\ll \sqrt{E_d}$ is a consequence of the fact that in 1D any weak interaction becomes strong at sufficiently low energies and the Born approximation (which is our first iteration) necessarily fails. To illustrate this we note that Eq.~(\ref{Eqalpha}) at low energies ($k,k',q\ll\sqrt{E_d}$) reads
\begin{equation}\label{EqalphaLowEn}
(k^2-q^2)\alpha_k+2Q \int \alpha_{k'} dk'/2\pi=0,
\end{equation}
where 
\begin{equation}\label{Q}
Q=4\pi E_d^{3/2}\langle 0|V+\hat{M}'| 0\rangle = E_d^2(C-C'\sqrt{E_d})
\end{equation}
is a small characteristic momentum. Equation~(\ref{EqalphaLowEn}) is nothing else than the 1D Schr\"odinger equation in momentum space describing the scattering on a weak $\delta$-function potential. Its exact solution is
\begin{equation}\label{alphasmallk}
\alpha_k=2\pi \delta(k-q)-\frac{Q}{q+iQ}\frac{2q}{k^2-q^2-i0}.
\end{equation}
We see that $\alpha_k$ is appreciable only at small momenta $\sim q\ll \sqrt{E_d}$, which justifies Eq.~(\ref{EqalphaLowEn}). Note that solving Eq.~(\ref{EqalphaLowEn}) in the first Born approximation we would miss the term $iQ$ in the denominator of Eq.~(\ref{alphasmallk}), which would lead to the $1/q$-divergence, exactly as we observe in Eqs.~(\ref{t1result}) and (\ref{r1result}).  

Now substituting Eq.~(\ref{alphasmallk}) into Eq.~(\ref{fexpansion}) and calculating the residues of $f(p)$ at $p=\pm q$ we obtain the transmission and reflection amplitudes:
\begin{equation}\label{rt1smallE}
r(q)= t(q)/t^{(0)}(q)-1=-iQ /(q+iQ).
\end{equation}
Equations (\ref{EqalphaLowEn}-\ref{rt1smallE}) are valid for $q^2\ll E_d$. This range of collision energies has a large overlap with the interval of validity of Eqs.~(\ref{t1result}) and (\ref{r1result}), and they can be easily matched with the low-energy result (\ref{rt1smallE}). A universal result can be obtained simply by making the substitution $q\rightarrow q+iQ$ in the denominators of Eqs.~(\ref{t1result}) and (\ref{r1result}).

Equation~(\ref{rt1smallE}) demonstrates that the integrable limit $E_d\rightarrow 0$ ($Q\rightarrow 0$) and the limit $q\rightarrow 0$ do not commute. In the near-integrable case ($E_d\ll 1$) the reflection probability is of order 1 for collision energies $q^2\lesssim Q^2 \propto E_d^4$. Outside of this small region $|r(q)|^2\propto E_d^4/q^2$. Recalling that in the near-integrable case $E_d\approx a^2$, we observe a very strong decrease of the reflection probability with decreasing $|a|$. Namely, at a fixed collision energy $|r|^2\propto a^8$.

The even, $F_s(q)$, and odd, $F_p(q)$, scattering amplitudes are related to the transmission and reflection amplitudes by the equations
\begin{equation}\label{FsFp}
F_s(q)=[t(q)+r(q)-1]/2,\,F_p(q)=[t(q)-r(q)-1]/2,
\end{equation}
and thus can be calculated to the first order from Eqs.~(\ref{t1result}) and (\ref{r1result}) for any collision energy. The even, $a_{ad}$, and odd, $b_{ad}$, scattering lengths can be defined through the effective range expansions of $F_s$ and $F_p$ valid for $q^2\ll E_d$:
\begin{equation}\label{F_s}
F_s(q)=-\frac{1}{1+2ia_{ad}q/\sqrt{3}+...}
\end{equation}
and
\begin{equation}\label{F_p}
F_p(q)=-\frac{1}{\sqrt{3}i/(2b_{ad}q)+1-2i\xi_p q/\sqrt{3}+...},
\end{equation}
where the rescaling of $q$ is due to the fact that the atom-dimer scattering momentum in our units equals $2q/\sqrt{3}$, which is the Fourier conjugate of the atom-dimer distance $\sqrt{3}x/2$. In Eq.~(\ref{F_p}) we also introduce the odd-channel effective range $\xi_p$, so that the expansions orders of $F_p$ and $F_s$ match. The even scattering length equals (cf. \cite{MoraA})
\begin{equation}\label{aad}
a_{ad}=\frac{3}{2}E_d^{-1/2}-\frac{11\sqrt{3}C}{12} E_d+o(E_d^{3/2}),
\end{equation}
the odd scattering length and effective range are given by
\begin{equation}\label{bad}
\frac{1}{b_{ad}}=\frac{2C}{\sqrt{3}}E_d^2 -\frac{2C'}{\sqrt{3}}E_d^{5/2}+o(E_d^{5/2})=\frac{2}{\sqrt{3}}Q,
\end{equation}
and
\begin{equation}\label{xip}
\xi_p=-\frac{3}{2}E_d^{-1/2}-\frac{23\sqrt{3}C}{12} E_d+3\sqrt{3} C' E_d^{3/2}+o(E_d^{3/2}).
\end{equation}

Let us now say a few words about higher order terms. The second order correction to $\alpha_k$ reads
\begin{equation}\label{alpha2}
\alpha_k^{(2)}=\frac{1}{\lambda_k - \lambda_q}\int\frac{dk'}{2\pi}\frac{\langle k|V+\hat{M}'| k'\rangle \langle k'|V+\hat{M}'| q\rangle}{\lambda_{k'} - \lambda_q}.
\end{equation}
The integral in Eq.~(\ref{alpha2}) diverges at large $k'$ if one uses the formulas (\ref{Vmatrix}-\ref{Mprimematrix}). Investigating Eqs.~(\ref{V}-\ref{Mprime}) we see that the exact matrix elements start decaying much more rapidly than the approximate ones (\ref{Vmatrix}-\ref{Mprimematrix}) at momenta $\gtrsim 1$. Therefore, in order to estimate $\alpha_k^{(2)}$ we can introduce a cut-off at $k'\sim 1$ in Eq.~(\ref{alpha2}) and still use Eqs.~(\ref{Vmatrix}-\ref{Mprimematrix}). One can then directly show that the leading term in $\alpha_k^{(2)}$ is by a factor $\sim\sqrt{E_d}$ smaller than the $\hat{M}'$-contribution to $\alpha_k^{(1)}$. This justifies our keeping the operator $\hat{M}'$ when calculating the first order terms. 

\section{Reaction probabilities}
\label{sec:measurable}

We can now discuss the reflection, transmission, and break-up probabilities, $|r(q)|^2$, $|t(q)|^2$, and $P_b(q)$. The reflection probability is approximated by $|r^{(1)}(q)|^2$ ensuring the two leading terms:
\begin{equation}\label{ReflProb}
|r(q)|^2\approx C^2\frac{E_d^4}{q^2}\left(\frac{E_d+3q^2}{E_d+q^2/3}\right)^2\left(1-2\frac{C'}{C}\frac{3E_d+q^2}{3\sqrt{E_d}}\right).
\end{equation}
The break-up probability can be obtained from the equation $P_b(q)=1-|r(q)|^2-|t(q)|^2$, in which $|t(q)|^2$ is calculated to the same order as $|r(q)|^2$ in Eq.~(\ref{ReflProb}):
\begin{equation}\label{|t|^2}
P_b(q)\approx -|r^{(1)}(q)|^2 -|t^{(1)}(q)|^2 - 2{\rm Re}\frac{t^{(1)}(q)+t^{(2)}(q)}{t^{(0)}(q)}.
\end{equation}
Here $t^{(2)}(q)$ is the correction to the transmission amplitude derived from Eq.~(\ref{alpha2}) in the same manner as $t^{(1)}(q)$ is derived from Eq.~(\ref{alpha1}):
\begin{equation}\label{t2}
\frac{t^{(2)}}{t^{(0)}}\!=\!\frac{4\pi i (q^2-E)^{3/2}}{q}\!\int\!\frac{dk'}{2\pi}\frac{\langle q|V+\hat{M}'| k'\rangle \langle k'|V+\hat{M}'| q\rangle}{\lambda_{q} - \lambda_{k'}}.
\end{equation}
From Eq.~(\ref{t1result}) one can see that ${\rm Re}[t^{(1)}(q)/t^{(0)}(q)]\equiv 0$. In fact, if $P_b(q)$ contained first order terms, we could play with the sign of the corresponding perturbation operator and make the probability negative, which is not possible in principle. Therefore, the operators that we have neglected earlier in favor of $V+\hat{M}'$ would contribute to Eq.~(\ref{|t|^2}) only to the second order and thus we do not exceed accuracy by keeping the second order term $t^{(2)}$. Anyway, the need for the second order terms here is formal as we know that the break-up probability can in principle be derived from the first order solution $\alpha_k^{(1)}$ by restoring the wavefunction $\Psi({\bf r}_1,{\bf r}_2)$ and calculating the outgoing flux of free atoms.

One can see that the real part of $t^{(2)}/t^{(0)}$ in Eq.~(\ref{t2}) originates only from the integration interval $k'\in (-\sqrt{E},\sqrt{E})$ and from the residues at $k'=\pm q$. The contribution of the latter cancels the first two terms in the right hand side of Eq.~(\ref{|t|^2}). This means that $P_b(q)\equiv 0$ below the break-up threshold as it should. For $E>0$ we have explicitly
\begin{equation}\label{P_bexplicit}
P_b(q)=-\frac{4 E_d^{3/2}}{q}\int_{-\sqrt{E}}^{\sqrt{E}}{\rm Im}\frac{\langle q|V+\hat{M}'| k'\rangle \langle k'|V+\hat{M}'| q\rangle}{\lambda_{k'} - \lambda_q}dk',
\end{equation}
and performing the integration we arrive at
\begin{equation}\label{P_bresult}
P_b(q)\approx \frac{C^2E_d^3(q^2-E_d)^3}{q^2(E_d+q^2/3)^2}\left[1-2\frac{C'}{C}\frac{E_d^2+10E_dq^2/3+q^4}{(\sqrt{E_d}+|q|)^3}\right].
\end{equation}
Just above the break-up threshold $P_b\propto E^3$. This is due to the fermionization of unbound states at low energies -- the probability to find three unbound interacting atoms close to each other scales with $E^3$ as in the case of three identical fermions.

Quantitative comparison of Eqs.~(\ref{ReflProb}) and (\ref{P_bresult}) shows that $|r|^2$ is significantly larger than $P_b$ even well above the break-up threshold (in the extreme limit $E_d\rightarrow 0$ the two curves intersect at $q^2/E_d \approx 12$), and, for example, for the collision energy $q^2=2 E_d$ the ratio $|r|^2/P_b \sim 40$ for $E_d\lesssim 0.1$. This means that the dissipative dynamics of a dimer immersed in a gas of atoms is dominated by the reflection rather than the break-up. On the other hand, the latter is a chemical reaction and its rate can be measured by monitoring the evolution of the population of dimers. The same holds for the break-up's inverse, the three-body recombination. The three-body recombination rate determines, for instance, the rate of dimer formation in a super-Tonks state -- the state of the system with negative coupling constant but without dimers \cite{McGuireAttractive,Astrakharchik2005,Batchelor} (the bosonic super-Tonks gas has been recently observed \cite{NagerlsuperTonks}).

The three-body recombination rate constant $\alpha_{\uparrow\uparrow\downarrow}(E)$ is readily obtained from $P_b(q)$ by using the principle of detailed balance. Indeed, the three-body recombination rate for a single $\uparrow$-$\uparrow$-$\downarrow$ triple per unit length in the center of mass reference frame equals $2\alpha_{\uparrow\uparrow\downarrow}(E)$. Here the factor $2$ comes from the fact that the product of densities $n_\uparrow^2 n_\downarrow$ is twice the number of $\uparrow$-$\uparrow$-$\downarrow$ triples per unit length. This rate multiplied by the density of unbound states (of a single triple) $\rho_{\uparrow\uparrow\downarrow}(E)=1/(4\sqrt{3}\pi)$ should equal the break-up rate $P_b(q)\sqrt{3}q$ multiplied by the density of atom-dimer states $\rho_{ad}(E)=1/(\sqrt{3}\pi q)$, where $E=-E_d+q^2$. In this manner we get
\begin{equation}\label{RecRate}
\alpha_{\uparrow\uparrow\downarrow}(E)=\alpha_{\uparrow\downarrow\downarrow}(E)=2\sqrt{3}P_b(q),
\end{equation}
which should be multiplied by $\hbar/m$ in order to restore the dimensions. We should mention that the rate constant (\ref{RecRate}) is averaged over a uniform distribution of the initial three-body unbound states in a small energy interval close to $E$, i.e. in the ergodic approximation. For highly nonthermal distributions one should speak about the differential recombination rate of a particular initial unbound state characterized by the asymptotic momenta (rapidities) and an auxiliary index related to the parametrization of the $\uparrow$-$\uparrow$-$\downarrow$ wavefunction in the nested Bethe ansatz picture \cite{Yang}. For each such state one can determine the zeroth order function $f^{(0)}$, expand it in the basis of $\chi_k$, substitute the corresponding coefficients $\alpha^{(0)}_k$ into the right hand side of Eq.~(\ref{Eqalpha}), determine $\alpha^{(1)}_k$, and finally derive the outgoing atom-dimer flux from the residues of the poles of $f^{(1)}(p)$ at $p=\pm q$. However, Eq.~(\ref{RecRate}) holds for a thermal gas (in this case $E\sim T$) and is also useful for systems in which the atomic momentum distribution is not extremely exotic. In particular, we believe that Eq.~(\ref{RecRate}) with $E\sim E_F$ gives a good estimate of the recombination rate constant in a degenerate super-Tonks gas, provided the Fermi energy $E_F=\pi^2n^2/2$ is much smaller than the dimer binding energy $E_d$.

Remarkable is that in the case $E\ll E_d$ the recombination rate constant (\ref{RecRate}) is independent of $E_d$, i.e. independent of the interatomic interaction strength. The reason for this is the following. On the one hand we have a nonintegrable perturbation, which acts on three atoms when they are close to each other. The squared modulus of this perturbation (relative to the zeroth order terms) scales as $E_d^3$ and it does vanish in the integrable case. On the other hand the local three-body correlation function (probability to find three atoms close together) is proportional to $(E/E_d)^3$. We see that when we multiply these two factors the $E_d$-dependence drops out. In a thermal gas we have $\alpha_{\uparrow\uparrow\downarrow}\propto (\hbar/m)(T/\hbar\omega)^3$.

\section{Bosonic case}
\label{sec:bosons} 

In the case of three identical quasi-1D bosons the derivations of Secs.~\ref{sec:STM} and \ref{sec:analytic} are essentially the same. The modifications are related only to the facts that the bosonic wavefunction is symmetric and that all three atoms interact with each other. Accordingly, the bosonic version of Eq.~(\ref{Thefequation}) differs from the fermionic one by an extra factor $-2$ in front of the operators $\hat{M}_{nm}$ \cite{MoraA}. The operators $\hat{M}'$ and $\hat{M}_{0m}$ in Eq.~(\ref{TheOperEq}) and the integral in Eq.~(\ref{L}) should also be multiplied by this factor. This changes the properties of the eigenstates and the structure of the spectrum of the operator $\hat{L}$ (we now have a trimer state separated from the continuum). 

In the bosonic case $\int \chi_q(p)dp$ is finite since three bosons can be at one point in space. This fact leads to important qualitative differences in between the fermionic and bosonic cases. Analyzing different terms in the bosonic version of Eq.~(\ref{TheOperEq}) in the same manner as we did for fermions in Sec.~\ref{sec:analytic} we see that the $M'$-term is the leading perturbation:
\begin{equation}\label{MprimeBosonic}
\hat{M}'f(p)=\frac{\log(4/3)}{2\sqrt{3}\pi}\int\frac{dk}{2\pi}f(k).
\end{equation}
It is of order $M'\sim \sqrt{E_d}$ when acting on the bosonic $\chi_q(p)$, whereas $V$ is still of order $E_d$ as in the fermionic case. One can also easily show that the last term in Eq.~(\ref{TheOperEq}) which represents the coupling to higher transverse states is of the same order of magnitude as the $V$-term. Therefore, the next to leading order correction is more difficult to obtain compared to the fermionic case. Here we restrict ourselves to the leading order correction originating from the $M'$-term, but before presenting the results let us briefly mention the effect of a narrow Feshbach resonance (cf. \cite{Yurovsky}). In this case we should add the term $\tilde{V}$ given by Eq.~(\ref{Vstar}) to the right hand side of Eq.~(\ref{TheOperEq}). We see, however, that it is larger than the $M'$-term only when the effective range $r_0$ is larger than the size of the 1D dimer $1/\sqrt{E_d}$, which is an extremely restrictive condition.

Formally, the $M'$-term, local in real space, can be understood as a modification of the one-dimensional interaction between two bosons in the presence of a nearby third boson. One can show that calculating the correction associated to this perturbation is equivalent to solving the Schr\"odinger equation
\begin{eqnarray}\label{BosonicSchrodinger}
&&\left\{-\nabla_{\{x_1,x_2\}}^2-2\sqrt{E_d}[\delta(x_1)+{\mbox{$\sum_\pm \delta(x_1/2\pm\sqrt{3}x_2/2)$}}]\right. \nonumber \\
&&\left.-8\sqrt{3}\log(4/3)E_d\delta(x_1)\delta(x_2)-E\right\}\Psi(x_1,x_2)=0
\end{eqnarray}
to the first order in the perturbation $-8\sqrt{3}\log(4/3)E_d\delta(x_1)\delta(x_2)$ (cf. \cite{Muryshev2002,Sinha,Mazets2008}). The unperturbed operator in the first line of Eq.~(\ref{BosonicSchrodinger}) is diagonalized by using the Bethe ansatz. The standard first order perturbation theory then gives the following results.

The first order correction to the boson-diboson transmission amplitude reads
\begin{equation}\label{t1resultBosons}
t^{(1)}(q)=i\frac{4\log(4/3)}{\sqrt{3}}\frac{E_d^{3/2}}{q}\frac{1+3q^2/E_d}{1+q^2/3E_d}t^{(0)}(q),
\end{equation}
where the zeroth order transmission amplitude equals
\begin{equation}\label{t0Bosons}
t^{(0)}(q)=\frac{1-\sqrt{3}iq/\sqrt{E_d}}{1+\sqrt{3}iq/\sqrt{E_d}}\frac{1-iq/\sqrt{3E_d}}{1+iq/\sqrt{3E_d}}.
\end{equation}
The zeroth order reflection amplitude vanishes and in the first order we get
\begin{equation}\label{r1resultBosons}
r^{(1)}(q)=i\frac{4\log(4/3)}{\sqrt{3}}\frac{E_d^{3/2}}{q}\left(\frac{1-\sqrt{3}iq/\sqrt{E_d}}{1+iq/\sqrt{3E_d}}\right)^2.
\end{equation}
Accordingly, to the leading order the reflection probability equals $|r^{(1)}|^2$. The break-up probability above the break-up threshold ($E=q^2-E_d>0$) is given by 
\begin{equation}\label{P_bresultBosons}
P_b(q)\approx \frac{16\log^2(4/3)}{3}\frac{E_d^{3/2}}{q^2}\frac{(q^2-E_d)^3}{(|q|+\sqrt{E_d})^3}\frac{1+3q^2/E_d}{1+q^2/3E_d}
\end{equation}
and the three-body recombination rate constant equals $\alpha_{rec}(E)=2\sqrt{3}P_b(q)$ as in the fermionic case. Here we also observe the scaling $P_b\propto\alpha_{rec}\propto E^3$ just above the threshold. This fact is related to the suppression of the local three-body density-density correlation function at low energies in an interacting 1D gas. That the two-body correlator is also suppressed can be seen by looking at the differential rate. Indeed, the recombination rate for three bosons on the length $L$ in a state parameterized by the set of rapidities $\{k_1,k_2,k_3\}$ equals
\begin{equation}\label{RecRateBosons}
\nu_{rec}(k_1,k_2,k_3)\approx 36\sqrt{3}\log^2(4/3)L^{-2}(E^3/E_d)\sin^2(3\phi),
\end{equation}
where $\phi=\arctan[\sqrt{3}(k_2-k_3)/(2k_1-k_2-k_3)]$ and we assume that the energy in the center of mass reference frame $E=\sum_i k_i^2/2-(\sum_i k_i)^2/6$ is much smaller than $E_d$. We observe that $\nu_{rec}\propto (k_i-k_j)^2$ for any pair of rapidities, if they are close to each other, $(k_i-k_j)^2\ll E$.

Due to the fact that the nonintegrable perturbation scales with lower power of $E_d$ compared to the fermionic case the reflection and break-up probabilities, as well as the rate of three-body recombination in the bosonic case contains one power of $E_d$ less. In particular, the three-body recombination rate actually increases with decreasing $E_d$ (although $E_d$ should be kept larger than $E$). This counterintuitive phenomenon is explained by the fact that the local three-body correlator is proportional to $(E/E_d)^3$, whereas the squared modulus of the perturbation is $\propto E_d^2$.

\section{Relaxation to deep states}
\label{sec:relaxation} 

Let us compare the reflection probability to the probability of relaxation to deeply bound molecular states. This inelastic process is extremely local. It takes place at distances $\sim R_e\ll 1$, where $R_e$ is the van der Waals range of the interatomic potential. The relaxation probability is thus proportional to the probability of finding three atoms at distances of order the oscillator length (unit in our case) multiplied by the recombination rate for three atoms confined to a unit 3D volume. We stay in the near-integrable regime where $|a|\ll 1$ and $E_d\approx a^2$. This gives the following results.

In the fermionic case the probability of relaxation in an atom-dimer collision at the collision energy $q^2$ equals
\begin{equation}\label{RelProb}
P_{rel}(q)\propto q^{-1}E_d^{9/2}(1+3q^2/E_d)(R_e/|a|)^{4+2\gamma},
\end{equation}
where $\gamma\approx-0.2273$ \cite{PSS}. Comparing Eqs.~(\ref{RelProb}) and (\ref{ReflProb}) we see that the relaxation probability is always much smaller than the reflection probability. One can also show that the recombination to deep molecular states is much slower than the formation of shallow dimers in a gas of unbound fermionic atoms. Namely, for $E\lesssim E_d$ the ratio of the corresponding recombination constants scales as $\alpha_{d,\uparrow\uparrow\downarrow}/\alpha_{\uparrow\uparrow\downarrow}\propto E_d(R_e/|a|)^{4+2\gamma}\ll 1$.

The relaxation probability in boson-diboson collisions can be calculated including the prefactor. Indeed, in the zeroth order the wavefunction of an atom and a dimer normalized per (axial) length $L$ is given by the axial Bethe ansatz wavefunction multiplied by the wavefunctions of the radial ground states. When the three atoms are at distances much smaller than 1 (the radial oscillator length) this wavefunction is approximately constant and its modulus square equals $|\Psi|^2=(1/9 \pi^2 L)\sqrt{E_d}(1+3q^2/E_d)/(1+q^2/3E_d)$. Here we use the normalization corresponding to the ``normal'' laboratory coordinates in which the atom-dimer distance is not rescaled (we mean the factor $\sqrt{3}/2$ in the definition of ${\bf r}_2$). The relaxation rate in this case equals $\nu_{rel}=\alpha_d(a<0)|\Psi|^2$, where $\alpha_d(a<0)$ is the relaxation rate constant for cold thermal bosons in the uniform space. This quantity has been calculated numerically \cite{Esry1999,BraatenHammer2001} and analytically \cite{MyLesHouches} in the zero-range theory as a function of $a$, the three-body parameter, and the elasticity parameter. We can write $\alpha_d(a<0)\approx C_d a^4$, where $C_d$ is a log-periodic function of $a$ bounded from below: $C_d\geq 128\pi^2(4\pi-3\sqrt{3})\coth(\pi s_0)\tanh\eta_*\approx 10^4\eta_*$. Here $s_0\approx 1.00624$ and the elasticity parameter is typically $\eta_*>0.1$. Assuming the case $C_d\approx 10^3$ we have
\begin{equation}\label{RelProbBoson}
P_{rel}(q)=(L/\sqrt{3}q)\nu_{rel}\approx 6.5 \frac{E_d^{5/2}}{q}\frac{1+3q^2/E_d}{1+q^2/3E_d}.
\end{equation}
We observe that the ratio $P_{rel}/|r|^2\gtrsim 15 q/\sqrt{E_d}$, which means that the relaxation is the dominant process in boson-diboson collisions unless we find a Feshbach resonance with a very low elasticity parameter and/or go to extremely low collision energies. 

The above analysis can also be performed for unbound states of three quasi-1D bosons. Namely, the relaxation rate of a Bethe ansatz state with rapidities $\{k_1,k_2,k_3\}$ equals $\nu_d(k_1,k_2,k_3)=\alpha_d(a<0)|\Psi|^2$, where $|\Psi|^2= (1/8\pi^2L^2)(E/E_d)^3\sin^2(3\phi)$. This rate scales as $1/E_d\approx 1/a^2$ in the same manner as the rate of three-body recombination to the weakly bound state (\ref{RecRateBosons}). In fact, these rates are related by $\nu_d/\nu_{rec}=2.5\times 10^{-3}C_d$ and we see that the recombination to deep states actually dominates, assuming, for example, the value $C_d=10^3$. 

\section{Discussion}
\label{sec:discussion}

A direct consequence of the relaxation analysis of the previous section is that for bosons, at least in the near-integrable case ($|a|\ll 1$), the thermalization due to the local three-body coupling of different Bethe ansatz states is likely to be slower than the relaxation and/or recombination to deeply bound molecules. This means that such a gas is stuck in a nonthermal state during all its lifetime (which may be long because of the 1D fermionization). In contrast, the relaxation processes in the fermionic case are suppressed and one should be able to observe the diffraction of momenta in atom-dimer collisions or the three-body recombination to a weakly bound state well before the gas decays. 

The dynamics of a mobile impurity in a 1D gas is at present a very attractive theme of theoretical (see \cite{Zvonarev,Gangardt} and references therein) and experimental \cite{Kohl,Inguscio} studies. The role of integrability in this problem has been discussed \cite{Colome} but is still far from being well understood. The present paper prepares grounds for studies in this direction. A quasi-1D $\uparrow\downarrow$ dimer immersed in a Fermi sea of $\uparrow$ species is a long-lived system in which, keeping the thermodynamic properties unchanged, one can modify the degree of integrability. Indeed, the thermodynamic quantities depend on the 1D $\uparrow$-$\downarrow$ coupling constant, which is proportional to the product of the scattering length $a$ and the frequency $\omega$ of the radial confinement \cite{Olshanii}. Keeping this product constant we can still modify the atom-dimer reflection probability in a wide range by using the fact that its scaling with $a$ and $\omega$ is drastically different. Namely, at a fixed collision energy we have $|r|^2\propto a^8\omega^5$ [see Eq.~(\ref{ReflProb})]. 

Having established control over the degree of integrability one can study various physical problems, e.g. whether or not a dimer looses its momentum while moving through a gas of atoms, i.e. whether or not there is a friction force on the dimer. Based on our understanding of the atom-dimer scattering problem we can conjecture what happens if one collides a gas of atoms with a gas of dimers in a quasi-1D trap with a weak axial parabolic confinement. In the integrable limit these two clouds pass through each other without reflection and one expects undamped relative oscillations. In contrast, as we increase the reflection probability these oscillations become damped and we expect to see none of them when $|r(q)|^2\gtrsim 1/N$, where $N$ is the particle number. This scenario remains based on the microscopic few-body analysis and we do not exclude nontrivial many-body effects, especially in the degenerate regime. We think that this problem deserves further experimental and theoretical investigation. 

Note that the limit $a\rightarrow 0$ does not at all correspond to the noninteracting case. Indeed, one can imagine a dimer in a state given by a wavepacket localized in momentum space around $k$ and in real space around $x_0+k t/2$. Let us assume that it passes an atom which is in a similar state but with $k=0$. This picture describes the relative motion of a dimer and an atom at high energies close to the bottom of the axial harmonic trap. After their collision the dimer wavepacket is centered around $x_0+\delta x +kt/2$, i.e. its trajectory is shifted by $\delta x$ (the atomic wavepacket is then centered at $-2\delta x$). By using the fermionic atom-dimer scattering solution (\ref{chi}) and the transmission amplitude (\ref{t0r0}) with $q=k/2\sqrt{3}$ one can show that this shift equals $\delta x=\sqrt{E_d}/(E_d+k^2/4)$. For the boson-diboson scattering we have $\delta x=\sqrt{E_d}[(E_d+k^2/4)^{-1}+(3E_d+k^2/12)^{-1}]$. Note that if $k^2$ is of order $E_d$, the shift is proportional to $1/\sqrt{E_d}\propto 1/|a|$ and actually increases with decreasing $|a|$. The dimer passes the atom faster than it does in the noninteracting case (as if the atom-dimer interaction is attractive). If we now assume that there is a gas of $N$ atoms with $N\gg 1$, their effect on the dimer's trajectory and on the frequency of its oscillations is appreciable. Note also that for a longitudinally trapped gas this shift is another source of nonintegrability even in the purely 1D case \cite{Mazets2011}.

As a more concrete experimental proposal for investigating the above phenomena we envision a setup, which is a combination of the fermionic experiment of Moritz {\it et al.} \cite{CI-molecule} on quasi-1D dimers and the bosonic quantum Newton's cradle experiment of Ref.~\cite{Kinoshita}. Namely, one can follow the routine of the former and create a quasi-1D spin mixture (in our case spin-imbalanced) of two different hyperfine states of $^{40}$K close to a zero crossing for the interspecies scattering length and form molecules by adiabatically decreasing the scattering length from zero to a finite $a<0$. One can then separate the dimers from atoms in momentum space by applying a Bragg pulse as has been demonstrated by Veeravalli {\it et al.} \cite{Vale} in the case of a 3D spin-mixture of $^6$Li. The transfer of $^{40}$K dimers into a state with momentum $4\pi/\lambda$, where $\lambda=767$~nm, can be done by using counterpropagating Bragg beams with the frequency detuning $\nu_{2m}=16.97$~kHz. Since the atoms remain at rest the atom-dimer scattering energy then equals $\nu_{2m}/3=5.66$~kHz. Assuming the radial confinement frequency $\nu_\perp=100$~kHz, which corresponds to the oscillator length $l_0=50$~nm, this sets $q^2=5.66\times 10^{-2}$ in our formulas. For $a=-25$~nm the dimer size equals $155$~nm, its binding energy $E_d\approx 10.5\times 10^{-2}$ (10.5~kHz in the laboratory units), and the reflection probability equals $|r(q)|^2=1.2\times 10^{-3}$. Under these conditions a cloud of, say, $30$ dimers will pass through a cloud of the same number of atoms several tens of times without thermalization. Increasing $a$ by a factor of 2 leads to $|r(q)|^2=0.01$ and we expect to see only very few oscillations. Although the value of $a$ in this case equals $l_0$, we show below that the binding energy calculated from Eq.~(\ref{BindingEnergy}), $E_d\approx 0.22$, is sufficiently small to apply our perturbation theory. 

\begin{figure}[hptb]
\begin{center}
\includegraphics[width=1\columnwidth,clip,angle=0]{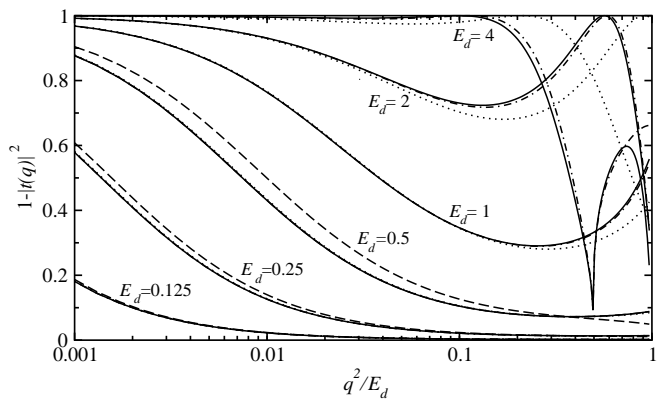}
\caption{The quantity $1-|t(q)|^2$ versus $q^2/E_d$ below the break-up threshold for several values of $E_d$. Numerical solutions of Eq.~(\ref{Thefequation}) are obtained by taking into account the first three (solid), the first two (dash-dotted), or only the ground (dotted) transverse states. The dashed lines is our perturbative result valid for small $E_d$. The dip at $q^2/E_d=0.5$ in the case $E_d=4$ is discussed in the text.}
\label{fig:ReflProb}
\end{center}
\end{figure}

For this type of experiments the relevant values of the reflection probability are small, of order $1/N$, meaning that we can use the perturbation theory valid for $E_d\ll 1$. In order to see the limitations of the perturbative approach we have also numerically solved Eq.~(\ref{Thefequation}) and calculated $|r(q)|^2$ and $|t(q)|^2$ for various $E_d$. In Fig.~\ref{fig:ReflProb} we show the quantity $1-|t(q)|^2$ versus $q^2/E_d$ below the break-up threshold for six different values of $E_d$ from 0.125 to 4. The perturbative result is shown as dashed lines and is given by Eq.~(\ref{ReflProb}) multiplied by $q^2/(q^2+Q^2)$, where $Q^2=C^2E_d^4(1-2C'\sqrt{E_d}/C)$ [see discussion after Eq.~(\ref{rt1smallE})]. We present it only for the three lowest values of $E_d$ since for larger $E_d$ the perturbation theory breaks down as expected -- the next-to-leading order term in Eq.~(\ref{ReflProb}) becomes comparable to the leading one.

For each value of $E_d$ in Fig.~\ref{fig:ReflProb} we also present three numerical curves: the solid lines are calculated by solving Eq.~(\ref{Thefequation}) projected to the first three transverse channels, i.e. we set $f_{n>2}\equiv 0$, the dash-dotted lines correspond to the first two channels, $f_{n>1}\equiv 0$, and the dotted lines are obtained by projecting Eq.~(\ref{Thefequation}) to the ground transverse state ($f_{n>0}\equiv 0$). For $E_d\lesssim 0.5$ the three lines practically coincide. For larger $E_d$ the projection to the ground transverse state is insufficient, the dotted lines are far off. However, we clearly observe a fast convergence with increasing the number of kept transverse states.

In the case $E_d=4$ the point $q^2/E_d=0.5$ ($q^2=2$) is the threshold for the excitation of the relative atom-dimer motion to the first transversely excited state (see the end of Sec.~\ref{sec:STM}). The corresponding branch-cut singularities are visible only in the solid and dash-dotted lines since the dotted line ignores all transverse channels other than the ground. In Fig.~\ref{fig:ReflProb} the difference in between the quantities $|r(q)|^2$ and $1-|t(q)|^2$ exists only in the case $E_d=4$ for $q^2/E_d>0.5$ where there is a finite probability for the atom-dimer pair to be excited into the first transverse channel.

The small-$q$ expansion of the transmission probability starts with the term $|t(q)|^2\approx (4/3)(a_{ad}+b_{ad})^2q^2$ which can be derived from Eqs.~(\ref{FsFp}-\ref{F_p}). In fact, one can show that the coefficient in front of $q^4$ is also proportional to $a_{ad}+b_{ad}$. Therefore, the point at which $a_{ad}=-b_{ad}$ is rather peculiar. In this case there is a large region of collision energies where atoms and dimers can be considered impenetrable. Then a dimer immersed in a gas of atoms can move only by shoving the atoms on its way, which leads to its diverging effective mass, subdiffusive propagation dynamics, etc. (see, for example, \cite{Zvonarev}). This regime is opposite to the integrable limit where we have no reflection. The condition $a_{ad}=-b_{ad}$ is reached for $E_d\approx 5.2$, but the effect is even more impressive at somewhat smaller $E_d$ (in Fig.~\ref{fig:ReflProb} we deliberately show the case $E_d=4$). Then the region of collision energies where the transmission probability is smaller than a certain small but finite value is wider.

Finally, let us comment on the dimer-dimer collisions. The scattering length for two quasi-1D dimers consisting of fermionic atoms has been calculated in Ref.~\cite{Mora4}. In principle, one can also find the dimer-dimer phase shift at finite collision energies. However, below the break-up threshold the two-dimer collisions do not lead to momentum diffusion even if the scattering phase shift differs significantly from the zeroth order one. This is because these are identical particles of the same mass. The momentum diffusion should appear in three-dimer collisions or in other processes involving more particles (atoms and/or dimers). Of course, these are much more difficult to analyze, but we can neglect these few-dimer diffusion channels compared to the two-body atom-dimer channel when the density of dimers is small or when they are in the Tonks regime, i.e. when the probability to find two dimers close together is suppressed.

\section{Summary and conclusion}
\label{sec:conclusions}

We have developed a perturbation theory for the quasi-1D three-atom problem in the near-integrable limit. We have shown that to the leading order the integrability of the $\uparrow$-$\uparrow$-$\downarrow$ fermionic system is broken by the effective range corrections to the two-body coupling constant originating from the virtual transverse excitations. In contrast, the quasi-1D problem of three bosons can be reduced to the purely 1D problem by adding an additional local three-body term \cite{Muryshev2002}, the two-body effective range corrections and other effects being of higher order. We have calculated the atom-dimer reflection, transmission, and break-up probabilities, as well as the three-body recombination rate constants for fermions and bosons as functions of the energy and the interatomic scattering length. At a finite collision energy $q^2$ the reflection probability is proportional to $a^8/l_0^{10}q^2$ in the fermionic case and to $a^6/l_0^8q^2$ for bosons.

We have shown that for fermions the rates of relaxation and recombination to deep molecular states are much lower than the rates of (momentum) diffusion in the zeroth order Bethe ansatz basis, meaning that the integrability breaking processes occur well within the lifetime of the system. In this respect the fermionic case differs strongly from the bosonic one. For bosons the rate of inelastic relaxation/recombination is comparable or higher than the rates of momentum diffusion. However, the lifetime of the bosonic gas is sufficiently long to study ``zeroth order'' effects of integrability, such as the reflectionless atom-dimer scattering. In spite of the absence of reflection the atom-dimer interaction is strong as can be seen from the phase of the transmission amplitude. The corresponding shift in position or in time of the relative atom-dimer trajectory can be measured, for example, in a quantum Newton's cradle type of experiment \cite{Kinoshita}.

The strong $a$-dependence of the atom-dimer reflection probability and the suppression of inelastic processes in a mixture of quasi-1D $\uparrow$-fermions and $\uparrow$-$\downarrow$-dimers makes this system an ideal candidate for a controllable investigation of differences in integrable and nonintegrable dynamics. We have estimated parameters of a realistic experiment, which can be performed in a spin-imbalanced mixture of two hyperfine states of $^{40}$K.

\begin{acknowledgments}
We thank N.~J. van Druten and M.~B. Zvonarev for fruitful discussions. DSP is supported by the Institut Francilien de Recherche sur les Atomes Froids (IFRAF) and by the Russian Foundation for Fundamental Research. This work is part of the research program on Quantum Gases of the Stichting voor Fundamenteel Onderzoek der Materie (FOM), which is financially supported by the Nederlandse Organisatie voor Wetenschappelijk Onderzoek (NWO).
\end{acknowledgments}


\end{document}